\documentclass[aps,prl,amsmath,amssymb,tightenlines,epsfig,floatfix,twocolumn,superscriptaddress, longbibliography]{revtex4-1}
\usepackage{graphicx}
\usepackage{dcolumn}	
\usepackage{bm}			
\usepackage{amsfonts}
\usepackage{xspace}
\usepackage{color}
\usepackage{epstopdf}
\usepackage{multirow}

\begin{document}

\newcommand{\psihat}{\ensuremath{\hat{\psi}}\xspace}
\newcommand{\psihatd}{\ensuremath{\hat{\psi}^{\dagger}}\xspace}
\newcommand{\ahat}{\ensuremath{\hat{a}}\xspace}
\newcommand{\Ham}{\ensuremath{\mathcal{H}}\xspace}
\newcommand{\ahatd}{\ensuremath{\hat{a}^{\dagger}}\xspace}
\newcommand{\bhat}{\ensuremath{\hat{b}}\xspace}
\newcommand{\bhatd}{\ensuremath{\hat{b}^{\dagger}}\xspace}
\newcommand{\boldr}{\ensuremath{\mathbf{r}}\xspace}
\newcommand{\dr}{\ensuremath{\,d^3\mathbf{r}}\xspace}
\newcommand{\tr}{\ensuremath{\,\mathrm{Tr}}\xspace}
\newcommand{\dk}{\ensuremath{\,d^3\mathbf{k}}\xspace}
\newcommand{\etal}{\emph{et al.\/}\xspace}
\newcommand{\ie}{i.e.\ }
\newcommand{\eq}[1]{Eq.\,(\ref{#1})\xspace}
\newcommand{\fig}[1]{Fig.\,\ref{#1}\xspace}
\newcommand{\abs}[1]{\left| #1 \right|}
\newcommand{\proj}[2]{\left| #1 \rangle\langle #2\right| \xspace}
\newcommand{\Qhat}{\ensuremath{\hat{Q}}\xspace}
\newcommand{\Qhatd}{\ensuremath{\hat{Q}^\dag}\xspace}
\newcommand{\phihatd}{\ensuremath{\hat{\phi}^{\dagger}}\xspace}
\newcommand{\phihat}{\ensuremath{\hat{\phi}}\xspace}
\newcommand{\boldk}{\ensuremath{\mathbf{k}}\xspace}
\newcommand{\boldp}{\ensuremath{\mathbf{p}}\xspace}
\newcommand{\boldsigma}{\ensuremath{\boldsymbol\sigma}\xspace}
\newcommand{\boldalpha}{\ensuremath{\boldsymbol\alpha}\xspace}
\newcommand{\grad}{\ensuremath{\boldsymbol\nabla}\xspace}
\newcommand{\parti}[2]{\frac{ \partial #1}{\partial #2} \xspace}
 \newcommand{\vs}[1]{\ensuremath{\boldsymbol{#1}}\xspace}
\renewcommand{\v}[1]{\ensuremath{\mathbf{#1}}\xspace}
\newcommand{\Psihat}{\ensuremath{\hat{\Psi}}\xspace}
\newcommand{\Psihatd}{\ensuremath{\hat{\Psi}^{\dagger}}\xspace}
\newcommand{\Vhatd}{\ensuremath{\hat{V}^{\dagger}}\xspace}
\newcommand{\Xhat}{\ensuremath{\hat{X}}\xspace}
\newcommand{\Xhatd}{\ensuremath{\hat{X}^{\dag}}\xspace}
\newcommand{\Yhat}{\ensuremath{\hat{Y}}\xspace}
\newcommand{\Jhat}{\ensuremath{\hat{J}}\xspace}
\newcommand{\Yhatd}{\ensuremath{\hat{Y}^{\dag}}\xspace}
\newcommand{\Uhat}{\ensuremath{\hat{U}^{\dag}}\xspace}
\newcommand{\jhat}{\ensuremath{\hat{J}}\xspace}
\newcommand{\lhat}{\ensuremath{\hat{L}}\xspace}
\newcommand{\Nhat}{\ensuremath{\hat{N}}\xspace}
\newcommand{\rhohat}{\ensuremath{\hat{\rho}}\xspace}
\newcommand{\ddt}{\ensuremath{\frac{d}{dt}}\xspace}
\newcommand{\nset}{\ensuremath{n_1, n_2,\dots, n_k}\xspace}
\newcommand{\Var}{\ensuremath{\mathrm{Var}}\xspace}
\newcommand{\Erf}{\ensuremath{\mathrm{Erf}}\xspace}
\newcommand{\tprep}{\ensuremath{\tau_\text{prep}}\xspace}
\newcommand{\tint}{\ensuremath{\tau_\text{int}}\xspace}

\newcommand{\notes}[1]{{\color{blue}#1}}
\newcommand{\sah}[1]{{\color{magenta}#1}}

\title{A Machine-Designed Sensor to Make Optimal Use of Entanglement-Generating Dynamics for Quantum Sensing}
\author{Simon A.~Haine}
\email{simon.a.haine@gmail.com}
\affiliation{Department of Quantum Science, Research School of Physics, Australian National University, Canberra, Australia}
\author{Joseph J.~Hope}
\affiliation{Department of Quantum Science, Research School of Physics, Australian National University, Canberra, Australia}

\begin{abstract}
We use machine optimisation to develop a quantum sensing scheme that achieves significantly better sensitivity than traditional schemes with the same quantum resources. Utilising one-axis twisting dynamics to generate quantum entanglement, we find that rather than dividing the temporal resources into seperate \emph{state-preparation} and \emph{interrogation} stages, a complicated machine-designed sequence of rotations allows for the generation of metrologically useful entanglement \emph{while} the parameter is interrogated. This provides much higher sensitivities for a given total time compared to states generated via traditional one-axis twisting schemes. This approach could be applied to other methods of generating quantum-enhanced states, allowing for atomic clocks, magnetometers, and inertial sensors with increased sensitivities. 
\end{abstract}

\maketitle

\noindent Atom interferometry is a crucial technique for enabling ultra-stable clocks, magnetometers, and inertial sensors \cite{Cronin:2009}. In the continued push for increased sensitivity of these devices, there is considerable recent interest in the development of atom interferometry that exploits quantum entanglement to surpass the shot-noise limit (SNL) \cite{Pezze:2016_review, Hosten:2016}. The large dimensionality of quantum Hilbert spaces means that even relatively simple quantum systems can display remarkably complicated dynamics. Physicists are very good at using intuition to find regimes that display simple behaviour within this complexity. Quantum sensors are usually designed accordingly, ensuring that the dynamics follows simple models, and the parameter of interest is robustly correlated with a simple output signal in a way that is easily interpreted.  These constraints do not necessarily maximise sensitivity, however.  If we relax these design constraints, it may be possible to use machine-based optimisation \cite{Rabitz:2004}  to design sensors with superior sensitivity and robustness. 

One-axis twisting (OAT) has been shown to produce spin-squeezed states capable of sub-SNL sensitivities \cite{Kitagawa:1993, Molmer:1999, Esteve:2008, Li:2009, Haine:2009, Leroux:2010, Gross:2010, Riedel:2010, Muessel:2014, Haine:2014, Nolan:2018}. However, one common criticism of this approach, or any approach that involves the preparation of quantum enhanced states, is that the time taken to prepare these states would be better utilised interrogating the parameter of interest. That is, instead of spending time $\tprep$ preparing an entangled quantum state, and time $\tint$ using this state to interrogate the system, it would be better to devote the full time period $T = \tprep + \tint$ to interrogation, thus increasing the sensitivity via simply accruing a greater phase-shift. Here, we consider how the use of state-preparation and interrogation \emph{concurrently} affects the performance of the sensor. In particular, we find that OAT dynamics, combined with a complicated sequence of rotations found by machine optimisation, can provide significantly better sensitivities than sensing with traditional schemes. 

\emph{Model ---} We assume a system of $N$ bosons distributed amongst two modes (annihilation operators $\ahat$ and $\bhat$).  Such a system is conveniently described by the pseudo-spin SU(2) algebra: $\jhat_k = \frac{1}{2}\bold{a}^\dag \sigma_k \bold{a}$, where $\bold{a} = (\ahat, \bhat)^T$, and $\sigma_k$ is the $k$th Pauli-spin matrix. These operators obey the usual angular momentum commutation relations: $\left[\jhat_i, \jhat_j\right] = i\sum_k \epsilon_{ijk} \jhat_k$, where $\epsilon_{ijk}$ is the Levi-Civita symbol \cite{Yurke:1986}. A general pure state can be expressed as $|\Psi\rangle = \sum_{m=-N/2}^{N/2} c_m |m\rangle$, where the state $|m\rangle \equiv |\frac{N}{2} +m, \frac{N}{2}- m\rangle$ denotes $\frac{N}{2}+m$ particles in mode $a$ and  $\frac{N}{2} -m$ particles in mode $b$, and is an eigenstate of $\jhat_z$ with eigenvalue $m$.  We assume that the metrological parameter, $\omega$ (for example, a frequency shift caused by a magnetic field) causes a rotation around the $\jhat_z$ axis and that the particles are interacting via a $\jhat_z^2$ interaction, the magnitude of which is constant in time. Additionally, we assume that we can implement an arbitrary time-dependent rotation around the $\jhat_x$ axis (rotation rate $\Omega(t)$), such that the Hamiltonian is 
\begin{equation}
\hat{H} = \hbar \chi \jhat_z^2 + \hbar \omega \jhat_z + \hbar \Omega(t)\jhat_x \, . \label{Ham}
\end{equation}
The $\jhat_z^2$ term is the source of the entanglement generation in OAT dynamics \cite{Toth:2012, Hyllus:2012, Hauke:2016}, and the presence of a constant, non-zero $\jhat_x$ term results in ``Twist-and-Turn" (TNT) dynamics, which has been shown to create entanglement more rapidly than OAT \cite{Micheli:2003, Strobel:2014, Muessel:2015, Sorelli:2019, Mirkhalaf:2018}. For an initial pure state $|\Psi_0\rangle$ evolving under this Hamiltonian for some duration $T$, the precision to which the parameter $\omega$ can be estimated by making measurements on the final state $|\Psi(T)\rangle$ is bounded by $(\Delta \omega)^2\geq \frac{1}{F_Q}$, where $F_Q$ is the quantum Fisher information (QFI) \cite{Toth:2014, Demkowicz-Dobrzanski:2014}, given by
\begin{align}
F_Q &= 4\left[\langle \partial_\omega \Psi(T)|\partial_\omega \Psi(T)\rangle - |\langle \Psi(T)| \partial_\omega \Psi(T)\rangle |^2\right] \, .
\end{align}
It was shown in \cite{Nolan:2017b} that as long as the initial state is an eigenstate of the $\jhat_x$ parity operator (such as a $\jhat_x$ eigenstate, for example) a measurement that projects into the $\jhat_x$ basis will saturate the quantum Cramer-Rao bound (QCRB), implying that we obtain the sensitivity $\Delta \omega^2 = 1/F_Q$. With $\Omega(t) = 0$, it is straightforward to show that 
\begin{align}
F_Q &= f_0 T^2 \, ,
\end{align}
where
\begin{equation}
f_0 = 4 \Var(\jhat_z)
\end{equation}
is the QFI for sensing an instantaneously encoded phase $\phi$ by making measurements on the state $|\Psi_\phi\rangle = \exp(i \phi \jhat_z)|\Psi \rangle$.  Assuming an initial state with no inter-particle entanglement (ie, a coherent spin state (CSS) \cite{Arecchi:1972}), the maximum possible value is $f_0 = N$, or $F_Q = NT^2$ \cite{Toth:2012, Hyllus:2012}, which we define as the shot-noise limit. In the absence of the $\jhat_x$ rotation, the entanglement generated by the $\jhat_z^2$ term does nothing to increase the sensitivity, as $\Var(\jhat_z)$, and therefore $f_0$, are conserved. In this letter we investigate the optimal form of $\Omega(t)$ to maximise the sensitivity at time $T$. We begin by optimising the traditional OAT scheme, where a single rotation is used to increase $f_0$ at some time, and then use machine optimisation to consider far more general forms of $\Omega(t)$. 

\begin{figure}
\centering 
\includegraphics[width =0.8\columnwidth]{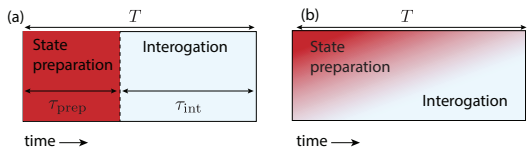}
\caption{(a): Traditional OAT scheme. With total available time $T$, time $\tprep$ is devoted to preparing the quantum state, while $\tint = T-\tprep$ is devoted to interrogating the system. (b): Machine-designed scheme customised to maximise the sensitivity, which includes dynamics which can't be separately classified as ``state-preparation" and ``interrogation". }
\label{timeline}
\end{figure}

\begin{figure}
\centering 
\includegraphics[width =1\columnwidth]{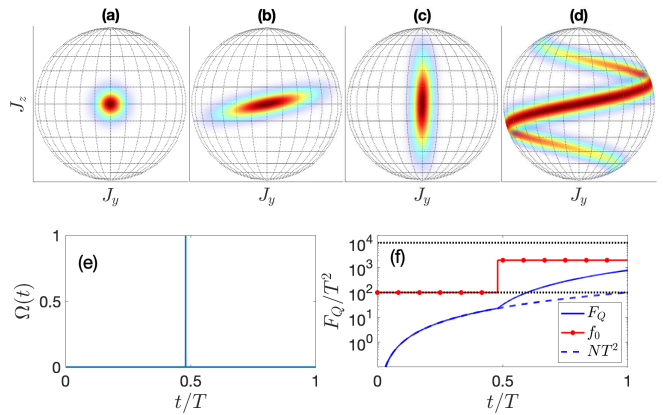}
\caption{Performance of the traditional OAT scheme. (a)-(d): The Husimi-$Q$ function, at (a): $t=0$, (b): $t=\tprep$ before the rotation is applied, (c) $t=\tprep$ \emph{after} the rotation $\exp(i \theta_0 \jhat_x)$ is applied, and (d): $t=T$. (e): $\Omega(t)$. In this case we chose $\Omega(t) = \theta_0 \delta(t-\tprep)$, resulting in an instantaneous rotation of magnitude $\theta_0$. (f): $F_Q(t)/T^2$ (blue solid line) and $f_0$ (red circles). $F_Q(t) = Nt^2$ for an un-entangled system is represented by the blue dashed line. The shot-noise limit ($f_0 = N$) and Heisenberg limit ($f_0 = N^2$) are represented by the lower and upper black dotted lines, respectively. Parameters: $N=100$, $\chi T = 0.1$. The parameters $\theta_0= -1.35$ and $\tprep = 0.48$ where chosen to maximise $F_Q(T)$.  The Husimi-$Q$ function is defined by $Q(\theta,\phi) = \abs{\langle \alpha(\theta,\phi)|\Psi\rangle}^2$, where $|\alpha(\theta,\phi)\rangle$ is a coherent spin state formed by rotating the maximal $\jhat_z$ eigenstate: $|\alpha(\theta,\phi)\rangle = \exp(i\phi \jhat_z)\exp(i\theta \jhat_y)|N/2\rangle$. }
\label{oat_scheme}
\end{figure}

\emph{Optimisation of traditional OAT scheme ---} OAT dynamics is usually discussed in the context of spin-squeezing \cite{Wineland:1992}, where the nonlinear interaction is used to create a state with reduced variance of the pseudo-spin operator along one direction. However, we will explain this process in the more general terms of QFI. In this protocol, at $t=0$ the state is prepared in a maximum $\jhat_x$ eigenstate: $|\Psi_0\rangle = \exp(i \frac{\pi}{2} \jhat_y)|N/2\rangle$, which evolves under \eq{Ham}. The $\jhat_z^2$ interaction causes a \emph{shearing} of the initial CSS, which increases $\Var(\jhat_y)$, while $\Var(\jhat_z)$ (and therefore $f_0$) are unaffected. In order to convert this entanglement into meteorologically useful entanglement (ie, increasing $f_0$), a rotation of angle $\theta_0$ is implemented around the $\jhat_x$ axis at $t=\tprep$ to convert this large $\Var(\jhat_y)$ into a large $f_0$, such that the sensitivity of the state to the parameter $\omega$ for the remaining time  $\tint = T-\tprep$ is significantly increased. This is done by choosing $\Omega(t)= \theta_0 \delta(t-\tprep)$, with $0< \tprep < T$ (\fig{timeline}) \footnote{In practice, $\Omega(t)$ is a very short temporal pulse with a characteristic timescale much shorter than the timescale for dynamics caused by the other terms in the Hamiltonian, with $\int \Omega(t) dt =\theta_0$}.   Although the parameter $\omega$ is being interrogated for the entire duration, $f_0$ significantly increases at $t=\tprep$, and then remains constant thereafter, so the notion of seperate state-preparation and interrogation periods remains useful. Figure \ref{oat_scheme} shows $f_0(t)$, $F_Q(t)$, and the evolution of the state generated via this scheme. The parameters $\tprep$ and $\theta_0$ were chosen to maximise $F_Q(T)$ for the particular value of $\chi T$. The use of OAT state-preparation dynamics concurrently with interrogation was also recently considered by Hayes \etal \cite{Hayes:2018} in a slightly different scheme. After time $\tprep$, the $\jhat_z^2$ term was switched off, and an \emph{echo} was performed by reversing the sign of the $\jhat_z^2$ term at time $T-\tprep$ to reverse the initial state preparation dynamics. We found that we could obtain significantly better sensitivity without this echo, and by optimising over $\theta_0$ \footnote{Hayes \etal used a slightly different Hamiltonian $\hat{H} = \hbar \chi \jhat_x^2 + \hbar \omega \jhat_y$, and therefore did not need this additional rotation. }. 

\begin{figure}
\centering 
\includegraphics[width =1\columnwidth]{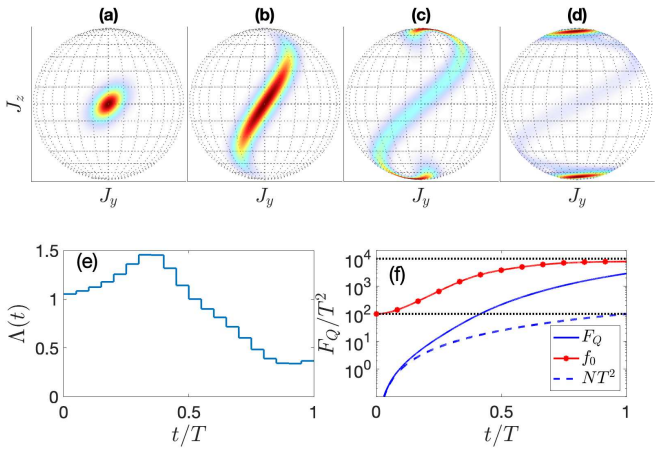}
\caption{Performance of the machine-designed scheme. (a)-(d): The Husimi-$Q$ function, at (a): $t=0.1T$, (b): $t=0.4T$, (c) $t=0.7T$, and (d): $t=T$. (e): $\Lambda(t)$. (f): $F_Q(t)/T^2$ (blue solid line) and $f_0$ (red circles). $F_Q(t) = Nt^2$ for an un-entangled system is represented by the blue dashed line. The shot-noise limit ($f_0 = N$) and  Heisenberg limit ($f_0 = N^2$) are represented by the lower and upper black dotted lines, respectively. Parameters: $N=100$, $\chi T = 0.1$. $F_Q$ reaches a maximum of $28.95 NT^2$, compared to $7.8 NT^2$ for the OAT scheme. }
\label{opt_scheme}
\end{figure}

\emph{Machine-designed scheme ---} The parameter $\Omega(t)$ can be manipulated with a high degree of control, for example, by adjusting the strength of an electromagnetic field. In traditional OAT and TNT metrology schemes, the complexity of dynamics afforded by a complicated form of $\Omega(t)$ has so far been neglected. We consider a more general class of dynamics by allowing for arbitrary choice of $\Omega(t)$ with the goal of maximising $F_Q(T)$. We note that this is different to maximising $f_0$, as rotations around $\jhat_x$, and subsequent dynamics, can partially over-write the phase accumulated at earlier times \cite{Haine:2018}. Thus, there is a trade-off between maximising the instantaneous value of $f_0(t)$, the rate of increase of $f_0(t)$, and preserving the phase accumulated at earlier times. To explore the large parameter space of possible temporal functions without restricting ourselves to solutions that correlate to simple intuitive models, we utilise machine optimisation. In situations where simple solutions are in fact optimal, these will arise organically.  We parameterise $\Omega(t)$ by
\begin{equation}
\Omega(t) = -\Lambda(t)N \frac{\chi}{2}
\end{equation}
where $\Lambda$ is a piece-wise step function.  Choosing a constant $\Lambda=1$ would correspond to TNT dynamics, but we divide the duration into 20 equal segments that are allowed to vary individually.  We then evolve our initial state under the Hamiltonian \eq{Ham} and use a gradient ascent algorithm with multiple, stochastically chosen starting locations to find the optimum $\Lambda(t)$ that maximises $F_Q(T)$. Figure \ref{opt_scheme} shows the optimum $\Lambda(t)$ for $\chi T = 0.1$ and $N = 100$. We find that $F_Q(T)$ is more than a factor of $3.5$ times better than the optimum OAT scheme, and $\sim 29$ times better than the shot-noise limit. 

\begin{figure}
\centering 
\includegraphics[width =1\columnwidth]{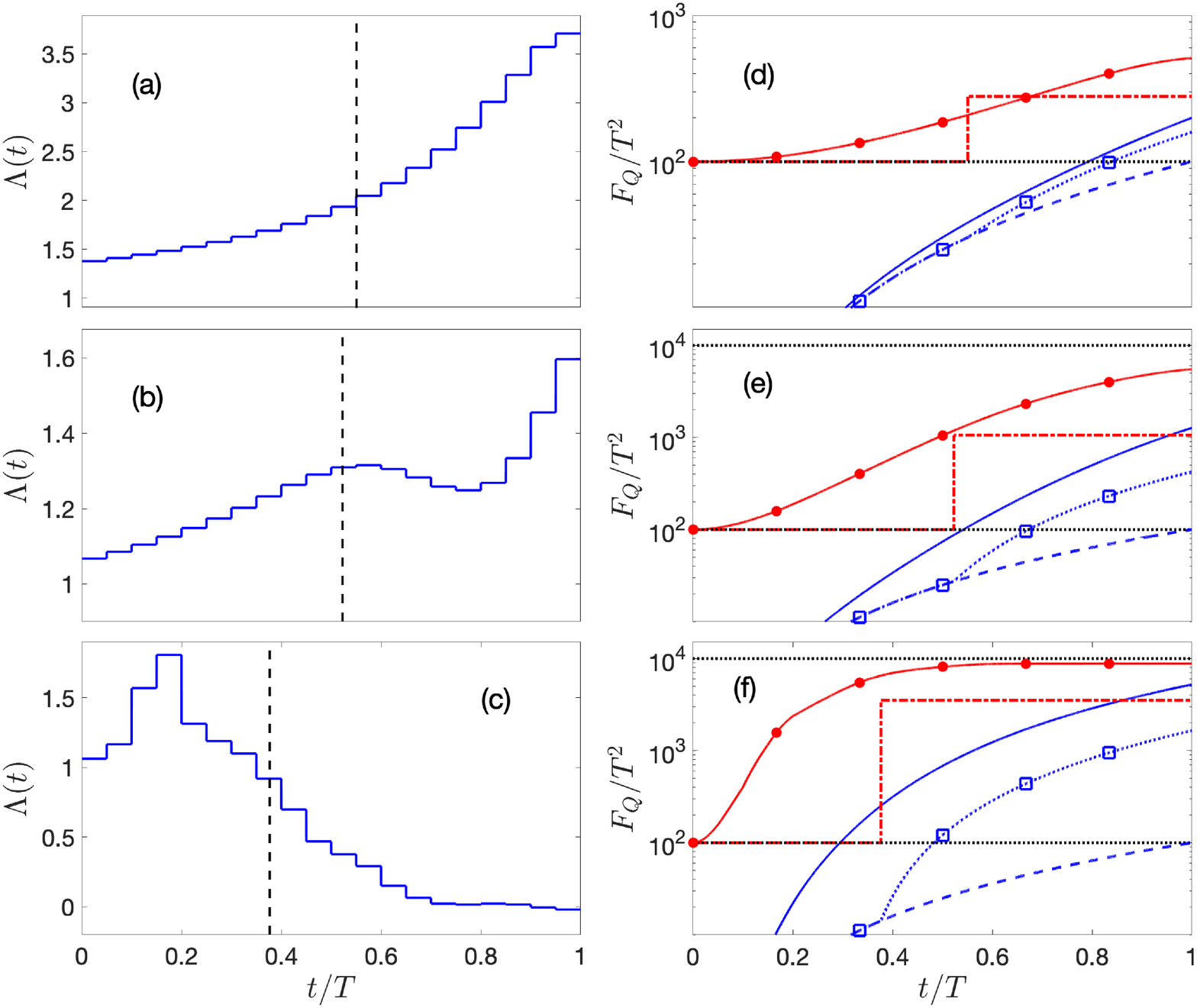}
\caption{Optimized scheme for different values of $\chi T$. (a)-(c): The optimum $\Lambda(t)$. The value of $\tprep$ for the optimum OAT scheme for the same parameters is indicated by the vertical dashed line. (d-f): $F_Q(t)$ and $f_0$ for our generalised scheme (blue solid line and red circles, respectively) and the optimum OAT scheme (blue squares and red dot-dashed line, respectively). The shot-noise limit ($F_Q = N T^2$ and $f_0 = N$) and Heisenberg limit ($f_0 = N^2$) are indicated by the blue dashed line, and the upper and lower black dotted lines, respectively. Parameters: (a,d): $\chi T = 0.02$, (b,e) $\chi T = 0.06$, (c,f): $\chi T = 0.2$. $N=100$ was used throughout.  }
\label{chi_compare}
\end{figure}

We examined the optimum scheme for a range for values of $\chi T$, and found different classes of behaviour in different regimes (\fig{chi_compare}). For $\chi T \lesssim 0.05 $, we find that the optimum behaviour seems to be continuously rotate the axis with increased variance into the $\jhat_z$ axis, which requires a larger $\Lambda$ as the state becomes more sheared. For $\chi T \gtrsim 0.09$, the optimum strategy seems to be more similar to what we would traditionally think of as seperate \emph{state preparation} and \emph{interrogation} stages,  which is to increase $f_0$ as much as possible early on (initially with $\Lambda \approx 1$ to initiate TNT dynamics), before increasing $\Lambda$ to rotate this sensitivity into the $\jhat_z$ axis, and then reducing $\Lambda(t) \approx 0$ for the remainder of the evolution. In all cases we find that using this more complicated profile for $\Lambda(t)$ significantly outperforms the traditional OAT and TNT schemes, even when $\tprep$ and $\theta_0$ are optimised to maximise the performance. 

\emph{Optimising Robustness to Detection Noise ---} To fully extract the sensitivity from these states, we require a measurement in the $\jhat_x$ basis with single-particle resolution \cite{Nolan:2017b}. However, in some practical situations this resolution is challenging, and there is additional detection noise. We follow the convention used in \cite{Pezze:2013} and model the behaviour of an imperfect detector by sampling from the probability distribution
\begin{equation}
\tilde{P}_m(\sigma) = \sum_{m^\prime} \Gamma_{m,m^\prime} P_{m^\prime} \, ,
\end{equation}
where
\begin{equation}
\Gamma_{m, m^\prime}(\sigma) =   e^{-(m-m^\prime)^2/(2\sigma^2)}/\sum_m e^{-(m-m^\prime)^2/(2\sigma^2)}
\end{equation}
convolves the raw probability distribution $P_m$ (i.e., the result of a $\jhat_x$ measurement with no detection noise) with a Gaussian of width $\sigma$. The sensitivity obtainable from sampling from $\tilde{P}_m$ is then $\delta \omega^2 = 1/\tilde{F}_C$, where 
\begin{equation}
\tilde{F}_C = \sum_m \frac{ \left(\partial_\omega \tilde{P}_m\right)^2}{\tilde{P}_m}
\end{equation}
is the classical Fisher information for this measurement. The states obtained by our optimised schemes are highly non-classical and are very susceptible to the effects of detection noise. In fact, for the state obtained in \fig{opt_scheme}, $\tilde{F}_C$ drops below the SNL for $\sigma \sim 1$.  

It was found in previous work that robustness to detection noise could be drastically improved by adding an interaction-based readout (IBR), which is a period of evolution after the interrogation time, to convert the final probability distribution into one that is more robust to detection noise \cite{Hosten:2016b, Frowis:2016, Nolan:2017b, Mirkhalaf:2018}. This often involves reversal of the initial state preparation dynamics (commonly referred to as an `echo') to restore the initial coherent spin-state \cite{Gabbrielli:2015, Davis:2016, Macri:2016, Linnemann:2016, Szigeti:2017, Fang:2017, Anders:2018, Huang:2018, Lewis-Swan:2018}. However, it was shown in \cite{Haine:2018b} that there are schemes that perform significantly better than this. Here, we consider the time devoted to this IBR as part of our total time $T$, and ask the question ``what is the optimum strategy in the presence of detection noise $\sigma$". Without dynamical control over the parameter $\chi$, we cannot implement a scheme that reverses the initial state preparation dynamics. However, it is possible that appropriate manipulation of $\Lambda(t)$ may approximate echo dynamics. We approach this by replacing $F_Q$ with $\tilde{F}_C$ as the metric in our optimisation algorithm. However, as in \cite{Nolan:2017b}, we found that the performance could be significantly improved by adding one additional parameter, which is a small phase offset before the final measurement. That is $|\Psi(T)\rangle \rightarrow\exp(i\phi \jhat_z)|\Psi(T)\rangle$, and optimize over the parameter $\phi$. 

\begin{figure}
\centering 
\includegraphics[width =1.0\columnwidth]{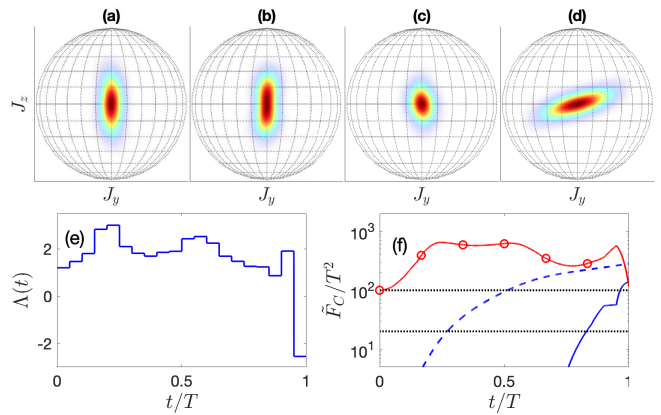}
\caption{Optimised scheme in the presense of detection noise. (a)-(d): The Husimi-$Q$ function, at (a): $t=0.25T$, (b): $t=0.5T$, (c) $t=0.75T$, and (d): $t=T$. (e): $\Lambda(t)$. (f): $\tilde{F}_C(t)/T^2$ (blue solid line), $F_Q(t)/T^2$ (blue dashed line) and $f_0$ (red circles). The SNL, and $\tilde{F}_C(T)$ for a CSS are indicated by the upper and lower black dotted lines, respectively. Parameters: $N=100$, $\chi T = 0.1$, $\sigma = 10$. $\tilde{F}_C$ reaches a maximum of $\sim 1.4 NT^2$. }
\label{noise_plot}
\end{figure}

Figure \ref{noise_plot} shows $\tilde{F}_C$ and the optimum shape of $\Lambda(t)$ for $\sigma=\sqrt{N}$, which represents a level of detection noise that usually reduces sensitivity to worse than the SNL \cite{Haine:2018b}. We see that while the early part of the scheme is still concerned with maximising $f_0$, the later part is attempting to restore the state to one which is less susceptible to detection noise.  The final value of $\tilde{F}_C \approx 1.4NT^2$ is considerably better than the SNL. For this value of $\sigma$, a CSS results in a sensitivity $\tilde{F}_C \approx 0.2 N T^2$. We note that instead of restoring to something approximating a CSS, the final distribution is anti-squeezed in $\jhat_x$, which was shown to provide greater robustness to detection noise \cite{Hosten:2016, Nolan:2017b, Haine:2018b}.

\begin{figure}
\centering 
\includegraphics[width =0.8\columnwidth]{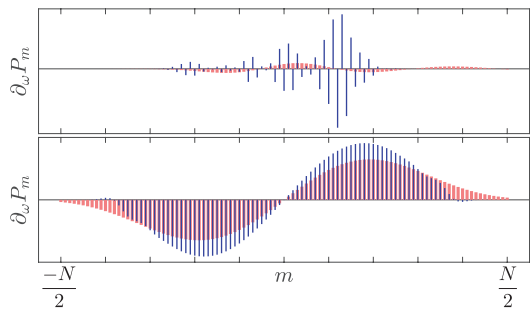}
\caption{Susceptibility of the final probability distribution to detection noise. $\partial_\omega P_m$ (narrow blue bars) and $\partial_\omega\tilde{P}_m$ (wide pink bars) for the scheme illustrated in figure \ref{opt_scheme} (top) and fig \ref{noise_plot} (bottom). In the top panel, $\partial_\omega\tilde{P}_m$ has been multiplied by $500$ in order for it to be displayed on the same scale. }
\label{prob_plot}
\end{figure}

By looking at the form of $\partial_\omega P_m$ and $\partial_\omega \tilde{P}_m$ (\fig{prob_plot}), we can see how this dynamics has increased the sensitivity in the presence of detection noise when compared to the scheme illustrated in figure \ref{opt_scheme}. In the scheme optimized without noise, the final state results in a probability distribution where $\partial_\omega P_m$ for neighbouring $m$ alternate in sign, so are washed out by detection noise of order $\sigma \sim 1$. When we optimise in the presence of detection noise, the scheme results in a  final probability distribution where a small change in $\omega$ simply results in a shift in the mean, which is only minimally effected by noise less than the width of the distribution. 

\emph{Discussion ---}  Hayes \etal \cite{Hayes:2018} also considered the use of traditional OAT dynamics concurrently with interrogation, and found that sensitivities significantly better than the SNL were possible. Comparing our scheme to Hayes's, for equivalent parameters, ($N=100$, $\chi T = 0.5$), we find a sensitivity $\sim7.2$ times better, while for $\chi T = 0.04$, our scheme achieves a sensitivity $\sim5.5$ times better than the SNL, while they find no improvement over the SNL.  Our scheme also outperforms TNT dynamics. 

The scheme presented in this paper is an example of a \emph{machine-designed sensor}, where instead of adhering to traditional sensing intuition (\ie seperate state-preparation and interrogation stages), we simply optimised the controllable parameters of the system using the final sensitivity as the appropriate metric. This approach frees us from the existing paradigm of state preparation followed by interrogation, which is the philosophy utilised by all quantum enhanced atomic sensing experiments so far demonstrated. The significant increase in performance that this approach provides indicates the power of this technique, and could be used in other quantum-enhanced sensing protocols that involve the use of a controllable dynamic parameter, such as when coherent coupling pulses are used to increase the entangled population spontaneously generated from spin-changing collisions \cite{Kruse:2016, Nolan:2016, Szigeti:2017}, four-wave mixing \cite{Haine:2011}, or Raman superradiance \cite{Haine:2013, Haine:2016}. Finally, we note that while this scheme is capable of enhancing the sensitivity of atomic clocks and magnetometers, the continuous use of coupling pulses is incompatible with atomic gravimeters and accelerometers due the the requirement of space-time separated modes \cite{Cronin:2009, Kritsotakis:2018}. However, this scheme could be useful in a Halkyard-Jones-Gardiner rotation sensor, due to the ability to continuously couple the constantly overlapping counter-rotating modes \cite{Halkyard:2010, Haine:2016b}. 
 
\begin{acknowledgements}
\emph{Acknowledgements---} We acknowledge fruitful discussions with Luca Pezze, Augusto Smerzi, Manuel Gessner, Jacob Dunningham, and Alex Radcliffe.   
\end{acknowledgements}

\bibliography{../../simon_bib.bib}

\end{document}